\documentclass[aps,pre,preprint,showpacs,superscriptaddress]{revtex4-1}
\usepackage{amsmath}
\usepackage{amssymb}
\usepackage{gensymb}
\usepackage{graphicx}
\usepackage{epstopdf}
\usepackage{soul}
\usepackage{setspace}
\usepackage{color}
\begin{document}
%\setstretch{1.5}

	\title{Direct reconstruction of the two-dimensional pair distribution function in systems with angular correlations}
		
	\author{I.~A.~Zaluzhnyy}
	\affiliation{Deutsches Elektronen-Synchrotron DESY, Notkestra{\ss}e 85, D-22607 Hamburg, Germany}
	\affiliation{National Research Nuclear University MEPhI (Moscow Engineering Physics Institute), Kashirskoe shosse 31, 115409 Moscow, Russia}
	
	\author{R.~P.~Kurta}
	\affiliation{European XFEL GmbH, Albert-Einstein-Ring 19, D-22761 Hamburg, Germany}	
	
	\author{A.~P.~Menushenkov}
	\affiliation{National Research Nuclear University MEPhI (Moscow Engineering Physics Institute), Kashirskoe shosse 31, 115409 Moscow, Russia}
	
	\author{B.~I.~Ostrovskii}
	\email[Corresponding author: ]{ostrenator@gmail.com}
	\affiliation{FSRC "Crystallography and Photonics", Russian Academy of Sciences, Leninskii prospect 59, 119333 Moscow, Russia}
	\affiliation{Landau Institute for Theoretical Physics, Russian Academy of Sciences, prospect akademika Semenova 1-A, 142432 Chernogolovka, Russia}
	
	\author{I.~A.~Vartanyants}
	\email[Corresponding author: ]{ivan.vartaniants@desy.de}
	\affiliation{Deutsches Elektronen-Synchrotron DESY, Notkestra{\ss}e 85, D-22607 Hamburg, Germany}
	\affiliation{National Research Nuclear University MEPhI (Moscow Engineering Physics Institute), Kashirskoe shosse 31, 115409 Moscow, Russia}

	\date{\today}
	
	\begin{abstract}
    An x-ray scattering approach to determine the two-dimensional (2D) pair distribution function (PDF) in partially ordered 2D systems is proposed.
    We derive relations between the structure factor and PDF that enable quantitative studies of positional and bond-orientational (BO) order in real space.
    We apply this approach in the x-ray study of a liquid crystal (LC) film undergoing the smectic-hexatic phase transition, to analyze the interplay between the positional and BO order during the temperature evolution of the LC film.
    We analyze the positional correlation length in different directions in real space.
%   By using the projection-slice theorem the correlation length corresponding to different directions in real space can be determined from the measured x-ray data.
%    We report on the experimental determination of the 2D PDF of a liquid crystal film undergoing the smectic-hexatic phase transition.
	\end{abstract}
	
	\maketitle
	
%\section{Introduction}

An absence of translational symmetry in disordered materials makes it challenging to characterize their structure and
establish a structure-functional relationship \cite{Barrat,Elliott1991, Cheng2011,Fraccia2015,Sellberg2014}.
Compared to crystalline matter, where a number of x-ray, neutron and electron scattering techniques are available
for structural characterization, much less information is accessible in the experiments on disordered and partially ordered materials \cite{Als-Nielsen, Chaikin}.
Despite the absence of long-range order present in crystals, disordered materials exhibit a number of structural features, such as short- and quasi-long-range order,
bond-orientational (BO) order, or dynamic heterogeneity, that can be also coupled to each other
%\cite{Steinhardt1981, Davey1984, Aeppli1984, Jeu2003,Kawasaki2007,Kapfer2015}.
\cite{Jeu2003,Kawasaki2007,Kapfer2015}.
Development of reliable characterization techniques capable of revealing various types of structural order is an important task in materials research.

The local structure of a system can be conveniently described by the pair-distribution function (PDF) $g(\textbf{r})$, which
defines the probability of finding a particle at the separation $\textbf{r}$ from any other arbitrary chosen particle \cite{Chaikin,Egami}.
The angular-averaged PDF $g(r)$, called also the radial distribution function, is typically used to characterize the structure of liquids and amorphous solids \cite{Als-Nielsen, Chaikin}.
The radial distribution function allows one to determine an average number of particles in a certain coordination shell and extract the positional correlation length.
However, this one-dimensional (1D) function is not sensitive to orientational order in the system,
that makes it difficult to use, for example, in the analysis of local atomic packing or BO order.

In this Letter we show that the two-dimensional (2D) PDF $g(\textbf{r})$ can be reconstructed directly from the measured diffraction patterns, that provides information on positional and orienational order in a partially ordered 2D system.
We applied this approach in the x-ray study of a liquid crystal (LC) film undergoing the smectic-hexatic phase transition, to analyze the interplay between the positional and BO order during the temperature evolution of the LC film.

%\section{Reconstruction of the 2D PDF}

\bigskip

Let us consider an x-ray scattering experiment on a homogeneous 2D system of $N$ identical particles,
where the direction of the incoming x-ray beam is perpendicular to the sample plane.
The structure factor $S(\textbf{q})$ of such a system is related to the
real-space PDF $g(\textbf{r})$ via the Fourier transform \cite{Als-Nielsen, Chaikin}
\begin{equation}
\label{EqIntensity}
S(\textbf{q})=I(\textbf{q})/N\lvert f(\textbf{q})\rvert^2=1+ \langle \rho \rangle \int{(g(\textbf{r})-1)e^{-i\textbf{q}\textbf{r}}\mathrm{d}\textbf{r}}\ ,
\end{equation}
where $I(\textbf{q})$ is the intensity measured in the forward scattering geometry at the momentum transfer vector $\textbf{q}$.
One can decompose both $S(\textbf{q})$ and $g(\textbf{r})$ into the angular Fourier series
\begin{equation}
\label{EqFourierG}
S(q,\phi)=\sum\limits_{n=-\infty}^{+\infty}S_n(q)e^{in\phi}, \quad
g(r,\theta)=\sum\limits_{n=-\infty}^{+\infty}g_n(r)e^{in\theta}\ ,
\end{equation}
where $\textbf{q}=(q,\phi)$, $\textbf{r}=(r,\theta)$ are the polar coordinates, and $S_n(q)$, $g_n(r)$ are
Fourier coefficients (FCs) of $S(q,\phi)$ and $g(r,\theta)$, respectively.
Then, by substituting Eqs.~(\ref{EqFourierG}) into (\ref{EqIntensity}), one can find
that FCs of the PDF $g_n(r)$ are related to FCs of the structure factor $S_n(q)$ via the Hankel transform (see Appendix \ref{AppA})
\begin{equation}
\label{EqGnToIn}
g_n(r)=\delta_{0,n}+\frac{1}{2\pi \langle \rho \rangle i^n }\int\limits_{0}^{+\infty}(S_n(q)-\delta_{0,n})J_n(qr)q\mathrm{d}q,
\end{equation}
where $\delta_{0,n}$ is the Kronecker delta and $J_n(qr)$ is the Bessel function of the first kind of integer order $n$.
By substituting the FCs (\ref{EqGnToIn}) into the Fourier series (\ref{EqFourierG}) one can readily determine the 2D PDF $g(\textbf{r})$ in real space.
If the form factor of the particles composing the system can be approximated to be isotropic $f(\textbf{q})=f(q)$,
then FCs of the structure factor $S_n(q)$ in Eq.~(\ref{EqGnToIn}) can be determined using FCs of the scattered intensity $I_n(q)$, $S_n(q)=I_n(q)/(N\lvert f(q)\rvert^2)$.
The latter can be calculated either directly from the measured diffraction patterns,
or utilizing the x-ray cross-correlation analysis (XCCA) \cite{Wochner2009,Altarelli2010,Kurta2012,Kurta2013a,Kurta2013}.
As it follows from this approach the 2D PDF can be fully determined by the corresponding experimentally measured 2D intensity distribution.

\begin{figure*}
	\includegraphics[width=\linewidth]{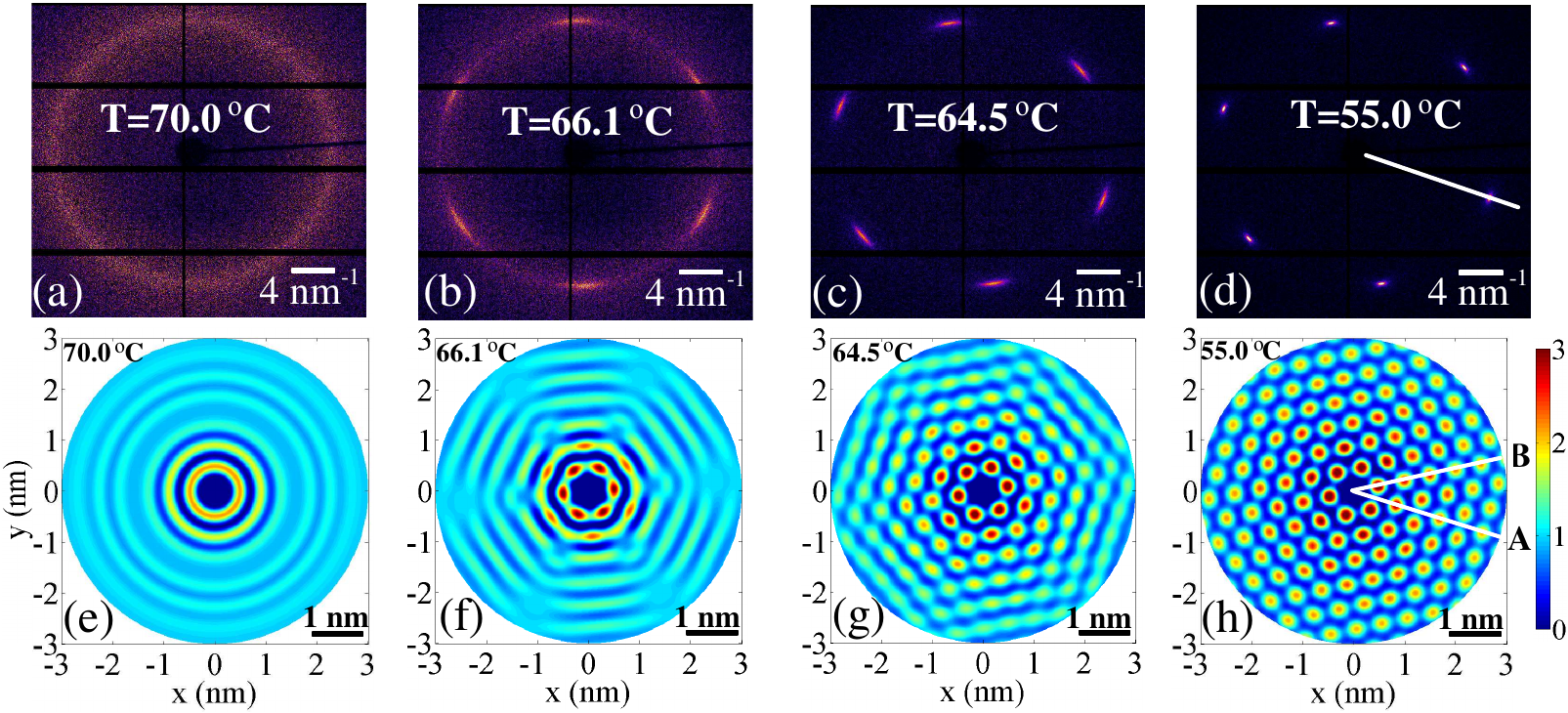}
	\caption{(a)-(d) Diffraction patterns measured at different temperatures from the LC film undergoing the smectic-hexatic phase transition.
		(e)-(h) The PDFs $g(\textbf{r})$ determined from the diffraction patterns (a)-(d), correspondingly.}
	\label{Detector}
\end{figure*}

In this Letter we applied the described approach to reconstruct the 2D PDF $g(\textbf{r})$ in a LC film undergoing the smectic-hexatic phase transition.
%In this letter we reconstructed the 2D PDF $g(\textbf{r})$ of a single LC layer in the smectic and hexatic phases.
An x-ray scattering experiment on a freely suspended LC film of the 3(10)OBC \cite{Huang1989} compound was conducted
at the beamline P10 of the PETRA III synchrotron source at DESY with photons of energy $E=13.6$ keV (see for experimental details Ref. \cite{Zaluzhnyy2015}).
The LC film consisted of about $2\cdot10^3$ layers that were formed by LC molecules oriented perpendicular to the layers.
An incident x-ray beam was perpendicular to the smectic layers and the detector was positioned downstream in the transmission geometry.
While decreasing the temperature of the LC film from $T=$70.0$~\degree$C to $T=$55.0$~\degree$C we observed
the smectic-hexatic phase transition at $T \approx 66.3~\degree$C \cite{Zaluzhnyy2015}.
Typical diffraction patterns measured in the smectic and hexatic phases are shown in Figs.~\ref{Detector}(a)-(d).
In the high-temperature smectic phase the measured diffraction pattern consists of a broad ring [Fig.~\ref{Detector}(a)],
while at lower temperatures in the hexatic phase the scattering ring splits into six arcs,
following the appearance and development of the BO order [Figs.~\ref{Detector}(b)-(d)].

The 2D PDFs, describing the structure of a single LC layer \cite{StructureFactorNote}, were determined at each temperature using Eqs.~(\ref{EqFourierG}) and (\ref{EqGnToIn}) and are shown in Figs.~\ref{Detector}(e)-(h).
The integration in Eq.~(\ref{EqGnToIn}) was performed up to the maximum value $q_{\text{max}}=1.9\;\text{\AA}^{-1}$ accessible with the detector at the given experimental conditions \cite{Zaluzhnyy2015}, that determined the real space resolution to be about  $3.3\;\text{\AA}$.
First, the form factor of a single 3(10)OBC molecule oriented along the x-ray beam was calculated numerically for the experimental conditions.
Then, due to fast rotation of the LC molecules around their long axis \cite{Seliger1978}, this form factor was averaged over all possible orientations of the molecule and can be considered to be isotropic within a layer.
In the frame of this approach the FCs of the structure factor can be directly obtained from the FCs of the scattered intensity using relations (\ref{EqIntensity}-\ref{EqFourierG}) \cite{FormFactorNote}.
%Due to the fast rotation of the LC molecules around their long axis \cite{Seliger1978}, the form factor $f(q)$ of a single molecule was considered to be isotropic within a layer \cite{FormFactorNote}.
%{More details of form factor evaluation}
In hexatic phase of LCs due to the sixfold symmetry of the diffraction patterns only FCs $S_n(q)$ of the order $n=6m$ ($m=0,1,2...$) have nonzero values.

At the temperature $T=70.0~\degree$C in the smectic phase the PDF represents a set of concentric circles [Fig.~\ref{Detector}(e)],
that corresponds to isotropic spatial distribution of molecules without angular correlations, which is typical for liquids and amorphous materials \cite{Als-Nielsen}.
%The rings have a width of about $3.5\;\text{\AA}$, that corresponds to a finite measured q-range.
The position of the first ring is $4.8\;\text{\AA}$, that corresponds to the average inter-molecular distance \cite{Jeu2003}. %$4\pi/q_0\sqrt{3}\approx5.1\;\text{\AA}$,
%where $q_0=1.41\;\text{\AA}^{-1}$ is the position of the maximum scattered intensity $I(q)$ \cite{Zaluzhnyy2015}.
%At large distances the value of $g(\textbf{r})$ approaches unity, which is an evidence of the absence of any positional correlations.
Slightly below the smectic-hexatic phase transition at $T=66.1~\degree$C the central uniform ring in the PDF splits into six well-separated bright spots [Fig.~\ref{Detector}(f)].
This sixfold angular modulation in the positions of the nearest-neighbor molecules appears due to emerging BO order that
induces angular anisotropy in the initially isotropic positional order.
One can also see that isotropy in particle correlations is also broken in the higher coordination shells,
where the concentric rings observed in the smectic phase transform to hexagons implying the presence of the hexatic phase.
At lower temperatures [Figs.~\ref{Detector}(g)-(h)] deeper in the hexatic phase, the tendency observed near the smectic-hexatic phase transition becomes more pronounced
due to increasing BO order.
At $T=55.0~\degree$C all molecular positions at small distances are well localized, and the PDF consists of separated sharp peaks [Fig.~\ref{Detector}(h)].
However one can notice that the peaks of
the PDF become more blurry at large values of r. This is a manifestation of the thermal angular fluctuations typical for hexatic films, which reveals itself at large distances.

%\section{Correlation length determination}

%
\begin{figure}
\includegraphics[width=\linewidth]{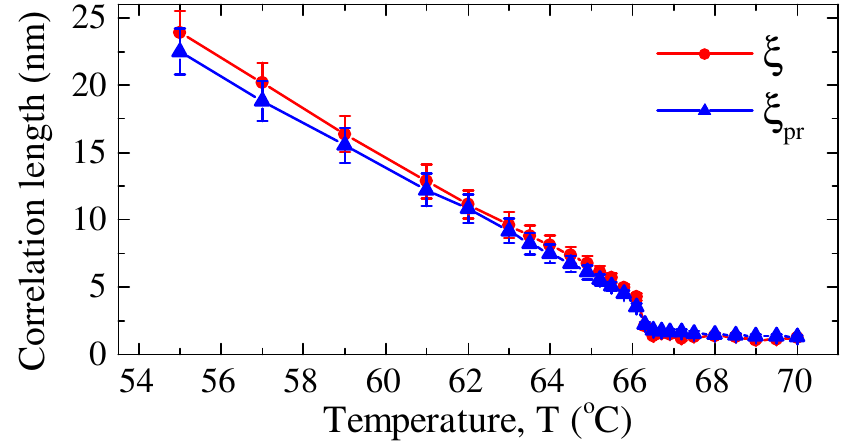}
\caption{Temperature dependence of the correlation length determined in different ways: $\xi$ from the radial section of scattered intensity (red circles) \cite{Zaluzhnyy2015} and $\xi_{\text{pr}}$ from the projection of the PDF $g(\textbf{r})$ onto the direction of the diffraction peak (blue triangles).
}
\label{CorrLength}
\end{figure}

\bigskip

One of the key components of the structural analysis of a partially ordered system is the positional correlation length $\xi$, which is a basic quantity to determine a length scale over which the positional correlations between the particles in the system decay.
%{\it Crystals are characterized by a long range order with the correlation length $\xi\rightarrow\infty$ short range order is described by the exponential decay of the PDF $~\exp(-r/{\xi}$...}}
Typically, it is evaluated as $\xi=1/\gamma$, where $\gamma$ is the half width at half maximum of the diffraction peak in the radial direction \cite{Stanley}.
In the previous work \cite{Zaluzhnyy2015} we determined the correlation length $\xi$ from the radial section of intensity $I(q)$ through a single diffraction peak in the direction shown with the white line in Fig.~\ref{Detector}(d).
The temperature dependence of the positional correlation length $\xi$ determined in such a way is presented in Fig.~\ref{CorrLength}.
It can be readily shown, by applying the projection-slice theorem \cite{ProjectSliceTheor}, that the same value of the correlation length can be also determined in real space from the projection of the function $g(\textbf{r})-1$ on the direction of a diffraction peak that is specified with white line $A$ in Fig.~\ref{Detector}(h).
%that the same value of the correlation length can be also determined in real space from the projection of the function $g(r)-1$ on the direction of a diffraction peak.
If the $x$-axis of Cartesian coordinates is chosen to be parallel to this direction, then the projection on this axis is (see Appendix \ref{AppC})
\begin{equation}
\label{EqPDFProjection}
g_{\text{pr}}(x)=\int_{-\infty}^{+\infty}\big(g(x,y)-1\big)\mathrm{d}y=A\cos{(q_0x)} e^{-\gamma |x|} \ ,
\end{equation}
where $A=2\pi\gamma/\langle\rho\rangle$.
The parameter $\gamma$ could be extracted from the exponential fit of the envelope function of the projected PDF $g_{\text{pr}}(x)$.

\begin{figure}
	\includegraphics[width=\linewidth]{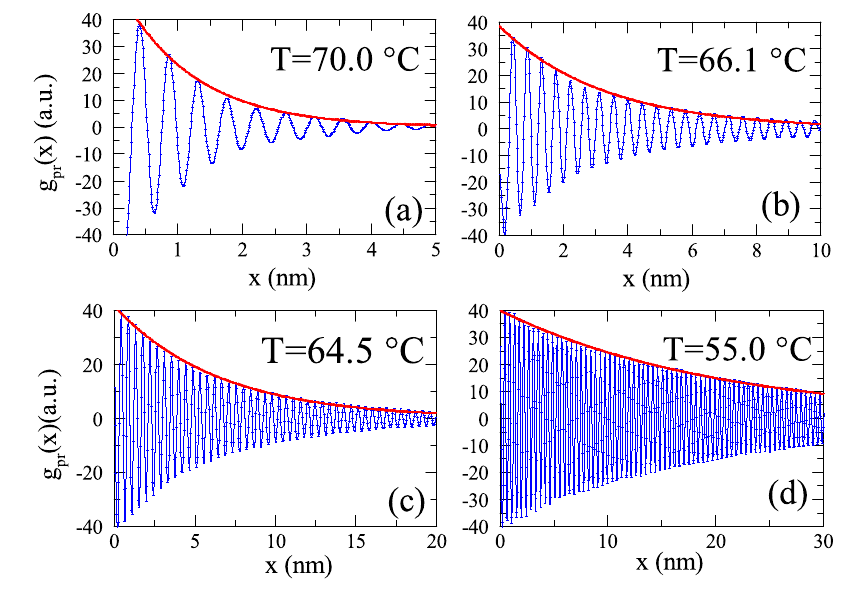}
	\caption{Projection of the PDF $g_{\text{pr}}(x)$ on the direction of the diffraction peak $A$ (see Fig.~\ref {Detector}(h)) at different temperatures.
		The projection $g_{\text{pr}}(x)$ is shown with the blue line and the envelope function in the form of an exponent $A\exp{(-\gamma x)}$ is shown with the red line.
	}
	\label{PDF_projection}
\end{figure}

The projection of the PDF $g_{\text{pr}}(x)$ on the direction $A$ %\textit{specified with the white line in Fig.~\ref{Detector}(h)}
is shown for different temperatures in Fig.~\ref{PDF_projection}.
As it is expected, $g_{\text{pr}}(x)$ is an oscillating function of distance with exponentially decreasing magnitude.
Since the peak position $q_0$ almost does not depend on temperature, the period of oscillations is the same for both the hexatic and smectic phase.
We determined the correlation length $\xi_{\text{pr}}=1/\gamma$ from the fit of the envelope function of $g_{\text{pr}}(x)$ in the form of an exponent $A\exp{(-\gamma x)}$
(see Eq. (\ref{EqPDFProjection})).
The obtained values of $\xi_{\text{pr}}$ as a function of temperature are shown in  Fig.~\ref{CorrLength}.
They are in a good agreement with the values of $\xi$ obtained from the radial scans of intensity $I(q)$.
The projection of the PDF on the direction between the diffraction peaks (direction $B$ in Fig.~\ref{Detector}(h)) is equal to zero in accordance with the projection-slice theorem, since there is no measured scattered signal in this direction.
%The projection of the PDF on the direction between the di raction peaks (shown with the red line in Fig.~\ref{Detector}(h)) is equal to zero in accordance with the projection-slice theorem, since there is no measured scattered signal in this direction.
We would like to note, if second order diffraction peaks could be measured in this direction this would lead to non-zero values of the corresponding PDF $g_{\text{pr}}(x)$.

\begin{figure}
	\includegraphics[width=\linewidth]{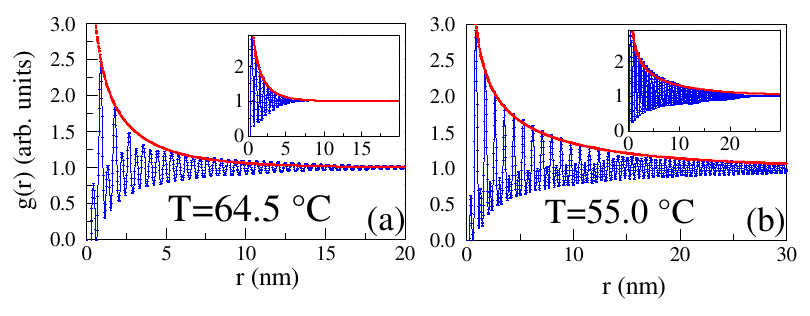}
	\caption{The radial section of the PDF $g(\textbf{r})$ in the direction of the diffraction peak $A$ (see Fig.~\ref {Detector}(h)) at different temperatures.
		In the insets the radial section of $g(\textbf{r})$ in the direction $B$ between the diffraction peaks (see Fig.~\ref{Detector}(h)) is shown at the same temperature.
	    The envelope function of the form $1+Br^{-0.5}\exp{(-\gamma r)}$, that was used for calculation of the correlation lengths $\xi_{\text{1,2}}=1/\gamma_{1,2}$, is shown with the red line.
	}
	\label{PDF_scan}
\end{figure}

Clearly, availability of the 2D PDF $g(\textbf{r})$ shown in Figs.~\ref{Detector}(e)-(h) allows us to analyze the intermolecular correlations in 2D real space in different directions.
%As an example, the radial sections of the PDF in two different directions indicated in Fig.~\ref{Detector}(h) with letters A and B at two different temperatures are shown in Fig.~\ref{PDF_scan} \cite{NegativePDF}.
As an example, the radial sections of the PDF in the directions $A$ and $B$, where index A corresponds to the direction through the diffraction peaks and index B - between the diffraction peaks (see Fig.~\ref{Detector}(h)) at two different temperatures are shown in Fig.~\ref{PDF_scan} \cite{NegativePDF}.
%In the smectic phase the diffraction pattern is isotropic, therefore two profiles of $g(\textbf{r})$ in different directions are identical [Fig.~\ref{PDF_scan}(a)].
Surprisingly, in the hexatic phase the PDF decays faster with a distance in the direction between the diffraction peaks (see insets in Figs.~\ref{PDF_scan}(a-b)).
We fitted the radial sections of the PDF shown in Fig.~\ref{PDF_scan} with an envelope of the form $1+Br^{-0.5}\exp{(-\gamma r)}$, where $B$ and $\gamma$ were the fitting parameters.
From these fits we determined the values of the positional correlation length $\xi_{A,B}=1/\gamma_{A,B}$ in two different directions $A$ and $B$ shown in Fig.~\ref{Detector}(h).
%where index A corresponds to the direction through the diffraction peaks and index B - between the diffraction peaks.
At the lowest measured temperature $T=55.0~\degree$C  in the hexatic phase we obtained values $\xi_A=17$ nm, $\xi_B=13$ nm as compared to the one $\xi_{pr}=23$ nm obtained from the projection-slice theorem.
We see from these results that the values of the correlation length strongly depend on the direction and are significantly lower than determined by the projection-slice theorem.
To understand this behavior we performed analysis in the frame of a general model described below.

We assume in the following that the structure factor of a monodomain system  can be represented as a product of two terms \cite{DifferentFCs}
\begin{equation}
\label{SFProduct}
S(q,\phi)=S(q)S(\phi),
\end{equation}
where $S(q)$ and $S(\phi)$  correspond to radial and angular dependence of the structure factor.
It is common to use the Lorentzian function to describe the radial profile of the structure factor $S(q)$ in a partially ordered material, $S(q)={\gamma^2}/{[(q-q_0)^2+\gamma^2]}$, where $q_0$ and $\gamma$ define the position and HWHM of a characteristic peak in the scattered intensity distribution \cite{Stanley}.
The angular part of the structure factor $S(\phi)$ can be quite generally represented as a Fourier series,
\begin{equation}
\label{AngularFourier}
S(\phi)=\sum\limits_{n=-\infty}^{+\infty}C_{n}e^{in\phi},
\end{equation}
where $C_{n}$ are FCs of $S(\phi)$, which can be defined to be real by an appropriate choice of the reference direction $\phi=0$.
For a monodomain system considered here, $C_{n}$ with $n\neq 0$ are nothing else but the generalized BO order parameters \cite{Steinhardt1981,Brock1989a}.
In the hexatic phase of a liquid crystal only parameters of the order $n=6,12,18,\ldots$ can have nonzero values,
due to sixfold rotational symmetry of the hexatic structure \cite{Brock1986}.
With Eqs. (\ref{SFProduct})-(\ref{AngularFourier}) one can readily describe diffraction peaks in the hexatic phase that are elongated in azimuthal direction as seen in Figs. \ref{Detector}(a)-(d).
%In general, if no orientational order is present in the system all $C_{n}$ in Eq.~(\ref{AngularFourier}) vanish, except a zero-order component $C_0$.
%In the opposite case of very strong BO ordering, non-zero coefficients  $C_{n}$ are positive and equal; then $S(\phi)$ becomes a set of sharp peaks, number of which is determined by the symmetry.

Using Eqs. (\ref{SFProduct}) and (\ref{AngularFourier}) together with the general formula (\ref{EqGnToIn}),
the following expression for the PDF can be obtained (see Appendix \ref{AppD})
\begin{equation}
\begin{split}
\label{PDFGeneral}
g(r,\theta) = 1+\frac{1}{2\pi\langle \rho \rangle} \sum\limits_{n=-\infty}^{+\infty}(-i)^n C_{n} e^{in\theta}\times \\
\times\int\limits_{0}^{+\infty}(S(q)-\delta_{0,n})J_{n}(qr)q\mathrm{d}q \, .
\end{split}
\end{equation}
It is interesting to note here that in Eq.~\eqref{SFProduct} for the structure factor the radial $q-$dependence and angular dependence were decoupled.
At the same time derived PDF in Eq.~\eqref{PDFGeneral} shows strong coupling between the positional and BO order.
This situation is typical for hexatic liquid crystals where the coupling between the density fluctuations and hexatic order parameters plays an important role \cite{Jeu2003}.
As a result both positional and BO order contribute to the PDF $g(r,\theta)$.

For large values of $r$ ($q_0r\gg n^2$) and small values of $\gamma$ ($\gamma\ll q_0$) the equation (\ref{PDFGeneral}) can be simplified to (see Appendix \ref{AppD})
\begin{equation}
\label{Asymptote}
g(r,\theta)\rightarrow 1+B\frac{e^{-\gamma r}}{\sqrt{r}}\cos \big(q_0r-\frac{\pi}{4} \big) \sum\limits_{n=-\infty}^{+\infty}C_{n}e^{in\theta},
\end{equation}
where $B=\gamma/\langle \rho \rangle \sqrt{q_0/2\pi}$.
At large distances the PDF $g(\textbf{r})$ exponentially decays and approaches unity due to an absence of any correlations at large values of $r$.
We note also that in asymptotic expression \eqref{Asymptote} the positional and BO order are decoupled.
The positional order is described by an exponentially decaying oscillating term and the BO order coincides with the angular dependence of the structure factor \eqref{AngularFourier}.
%The asymptotic decay of correlations is determined by the second term in (\ref{Asymptote}) that contains a product of two parts.
%The first has an oscillatory behavior as a function of distance with an exponentially decaying factor.
%The second part describes angular anisotropy of the PDF and thus coincides with the angular dependence of the structure factor (\ref{AngularFourier}).
In the hexatic phase the angular part of the structure factor $S(\phi)$ has a maximum in the direction of the peaks maximum and approaches zero between the diffraction peaks.
As such, the oscillations of the PDF $g(r,\theta)$ are suppressed in these directions (Fig. \ref{PDF_scan}).
Our analysis shows that in the frame of our model even asymptotically the positional correlation length can show an apparent different values in different directions.

%In the isotropic smectic phase $S(\phi)$ is constant, so all FCs $C_n$ are zero except $C_0$.
%Thus, the PDF $g(r,\theta)$ is also isotropic in the smectic phase (Fig. \ref{PDF_scan}(a)).
%This anisotropy would disappear, if we measured the diffraction peaks of higher orders.
%In accordance with Eq. (\ref{Asymptote}),

%Although the reconstructed PDF is well fitted by the chosen envelope function (see Fig. \ref{PDF_scan}), it appears that in the hexatic phase $\xi_2<\xi_1<\xi$.
%The fact that $\xi_2<\xi_1$ is due to suppression of the PDF oscillations in the direction between the diffraction peaks.
%If the measurement of the higher order diffraction peaks were possible, the angular anisotropy of the PDF at large distances would disappear and the values of the correlation length would converge.
%The second inequality ($\xi_1,\xi_2<\xi$) is due to the fact that Eq. (\ref{Asymptote}) is valid for large distances only; an attempt to fit the PDF with exponential envelope for all distances results in an underestimated value of the positional correlation length.

The situation will be different if measurements of the higher order diffraction peaks were possible.
In this case the angular anisotropy of the PDF at large distances would disappear and the values of the correlation length would converge.
Our analysis showed that the values of the correlation length in different directions will be also the same in the case of symmetrical Bragg peaks, i.e. not elongated in azimuthal direction.

%In the smectic phase, where there is no angular dependence of the structure factor, all three values of the positional correlation length coincide $\xi_2=\xi_1=\xi$.
%It is interesting to note, that in the case of symmetrical Bragg peaks, i.e. not elongated in azimuthal direction, the values of the correlation length $\xi_1$ and $\xi_2$ measured in two different directions would be equal {\it even when the BO order is present.???}
%The above analysis indicates that the interplay between BO and positional order in hexatic films under certain conditions can lead to an apparent anisotropy of the positional correlations.

%\section{Conclusions}

\bigskip

In summary, we proposed an approach for determination of the 2D pair distribution function in partially ordered 2D systems of identical particles with angular correlations directly from the experimentally measured x-ray diffraction patterns.
The derived relations between the structure factor and 2D PDF have been used to reconstruct the PDF in hexatic and smectic films of liquid crystals.
Application of the projection-slice theorem allowed to determine the values of the correlation length in systems at different temperatures.
Deduced PDF data clearly indicate both the exponential decay characteristic of the short-range positional in-plane order in
hexatics, and the long-range BO order in the arrangement of the maximums of the PDF on the 2D maps in real space.
By introducing a model describing scattering data from hexatic films we analyzed in details the behavior of the 2D PDF and explained an apparent different decay of the positional correlation length in different directions.
%
%Application of the project-slice theorem allowed the values of the correlation length corresponding to different directions in real space to be compared.
%The determined 2D PDFs show that spatial correlations between the LC molecules in the hexatic phase decay differently in distinct directions, indicating strong coupling of positional and bond-orientational order.
%The proposed approach is particularly advantageous for analysis of spatial anisotropies of inter-particle correlations.

We foresee that the proposed method of the 2D PDFs reconstruction can be widely used for analysis of partially ordered systems such as polymer colloids, suspensions of biological molecules, block copolymers and liquid crystals.
Our approach is particularly advantageous for analysis of spatial anisotropies of inter-particle correlations.

\bigskip

\begin{acknowledgments}
We acknowledge support of this project and discussions with E. Weckert.
%We are thankful to C.~C. Huang, W.~H. de Jeu, E.~I. Kats, V.~V. Lebedev, A.~R. Muratov, and V.~M. Kaganer for fruitful discussions and R.~Gehrke for careful reading of the paper.
We are thankful to A. Zozulya and M. Sprung as well as to the whole team of Coherence beamline P10 at synchrotron source PETRA III for their support during beamtime and to S. Funary for a careful reading of the manuscript.
This work was partially supported by the Virtual Institute VH-VI-403 of the Helmholtz Association.
The work of I.A.Z. and B.I.O. was partially supported by the Russian Science Foundation (grant 14-12-00475).
\end{acknowledgments}

\appendix\section{Reconstruction of the Pair Distribution Function}
\label{AppA}
The structure factor $S(\textbf{q})$ and pair distribution function (PDF) $g(\textbf{r})$ are related to each other by the Fourier transform \cite{Chaikin,Als-Nielsen}
\begin{equation}
\label{Eq1}
S(\textbf{q})=1+ \langle \rho \rangle \int{(g(\textbf{r})-1)e^{-i\textbf{q}\textbf{r}}\mathrm{d}\textbf{r}} \, ,
\end{equation}
where $\langle \rho \rangle$ is an average density of the particles and $\textbf{q}$ and $\textbf{r}$ are vectors in reciprocal and real space respectively.
In the case of a two-dimensional (2D) system it is convenient to use the polar coordinates, i.e. $\textbf{r}=\{r,\theta\}$ and $\textbf{q}=\{q,\phi\}$.
Since the function $g(\textbf{r})=g(r,\theta)$ is periodic over the angular variable $g(r,\theta+2\pi)=g(r,\theta)$, one can decompose it into a Fourier series
\begin{equation}
\label{Eq2}
g(r,\theta)=\sum\limits_{n=-\infty}^{+\infty}g_n(r)e^{in\theta} \, ,
\end{equation}
where $g_n(r)$ are the Fourier coefficients
\begin{equation}
\label{Eq3}
g_n(r)=\frac{1}{2\pi}\int\limits_{0}^{2\pi}g(r,\theta)e^{-in\theta}\mathrm{d}\theta\, .
\end{equation}
In case of absence of any correlations $g(\textbf{r})=1$ and $g_{n}(r)=\delta_{0,n}$, where $\delta_{0,n}$ is the Kroneker delta.
Similar Fourier series over the angular variable can be written for the structure factor
\begin{equation}
\label{Eq4}
S(q,\phi)=\sum\limits_{n=-\infty}^{+\infty}S_n(q)e^{in\phi} \, ,
\end{equation}
where the Fourier coefficients $S_n(q)$ are
\begin{equation}
\label{Eq5}
S_n(q)=\frac{1}{2\pi}\int\limits_{0}^{2\pi}S(q,\phi)e^{-in\phi}\mathrm{d}\phi \, .
\end{equation}

Substituting expansion (\ref{Eq2}) into the integral (\ref{Eq1}) and interchanging operations of integration and summation we obtain
\begin{equation}
\begin{split}
\label{Eq6}
S(q,\phi)= & 1+\langle \rho\rangle \int\limits_{0}^{2\pi}\int\limits_{0}^{+\infty}
\bigg( \sum\limits_{n=-\infty}^{+\infty}g_n(r) e^{in\theta}-1 \bigg) e^{i\textbf{qr}}r\mathrm{d}r\mathrm{d}\theta= \\
&= 1+\langle \rho\rangle \sum\limits_{n=-\infty}^{+\infty} \int\limits_{0}^{2\pi}\int\limits_{0}^{+\infty}
\big( g_n(r) -\delta_{0,n} \big) e^{in\theta}  e^{iqr\cos(\theta-\phi)}r\mathrm{d}r\mathrm{d}\theta= \\
&= 1+\langle \rho\rangle \sum\limits_{n=-\infty}^{+\infty} e^{in\phi} \int\limits_{0}^{+\infty}\mathrm{d}r\, r\big( g_n(r) -\delta_{0,n}\big)\int\limits_{0}^{2\pi}e^{in\Psi}e^{iqr\cos\Psi}\mathrm{d}\Psi \, ,
\end{split}
\end{equation}
where $\Psi=\theta-\phi$.
Now using the integral representation of the Bessel function \cite{Watson}
\begin{equation}
\label{Eq7}
J_n(qr)=\frac{1}{2\pi i^n}\int\limits_{0}^{2\pi}e^{in\Psi}e^{iqr\cos\Psi}\mathrm{d}\Psi \, ,
\end{equation}
we obtain for the structure factor
\begin{equation}
\label{Eq8}
S(q,\phi)=1+2\pi \langle \rho \rangle \sum\limits_{n=-\infty}^{+\infty} e^{in\phi}i^{n} \int\limits_{0}^{+\infty}\big( g_n(r) -\delta_{0,n}\big)J_n(qr)r\mathrm{d}r \, .
\end{equation}
%

%Substituting Fourier series (\ref{Eq4}) into Eq. (\ref{Eq8}) one obtains for the Fourier components of the structure factor
By comparing (\ref{Eq4}) and (\ref{Eq8}) we obtain for the Fourier components of the structure factor
\begin{equation}
\label{Eq11}
S_n(q)=\delta_{0,n}+2\pi i^n \langle \rho \rangle \int\limits_{0}^{+\infty}\big( g_n(r) -\delta_{0,n}\big)J_n(qr)r\mathrm{d}r \, .
\end{equation}
Thus, the Fourier coefficients of the structure factor $S_n(q)$ are related to the Fourier coefficients of the PDF $g_n(r)$ via the Hankel transform \cite{Poularikas}.
Rearranging Eq. (\ref{Eq11}), multiplying it by $qJ_n(qr')$, and integrating it over $q$ from 0 to infinity, one obtains
\begin{equation}
\label{Eq11a}
\frac{1}{2\pi i^n \langle \rho \rangle}\int\limits_{0}^{+\infty}\big(S_n(q)-\delta_{0,n}\big)J_n(qr')q\mathrm{d}q=\int\limits^{+\infty}_{0} \int\limits^{+\infty}_{0} \big(g_n(r)-\delta_{0,n}\big)J_n(qr)J_n(qr')r\mathrm{d}rq\mathrm{d}q \, .
\end{equation}
Using the orthogonality property of Bessel functions for $r,r'>0$
\begin{equation}
\label{Eq11b}
\int\limits_{0}^{+\infty}J_n(qr)J_n(qr')qdq=\frac{\delta(r-r')}{r}\, ,
\end{equation}
where $\delta(r)$ is the Dirac delta function, we can express Fourier coefficients of the PDF through Fourier coefficients of the structure factor
\begin{equation}
\label{Eq12}
g_n(r)=\delta_{0,n}+\frac{1}{2\pi \langle \rho \rangle i^n }\int\limits_{0}^{+\infty}(S_n(q)-\delta_{0,n})J_n(qr)q\mathrm{d}q\, .
\end{equation}

If the Fourier coefficients of the structure factor $S_n(q)$ can be determined from a diffraction experiment, the Fourier coefficients of the PDF $g_n(r)$ can be calculated using Eq. (\ref{Eq12}) and then the PDF $g(\textbf{r})$ can be reconstructed through the series (\ref{Eq2}).

\section{X-ray diffraction from a thick liquid crystal film}
\label{AppB}
\begin{figure*}
	\includegraphics[width=65mm]{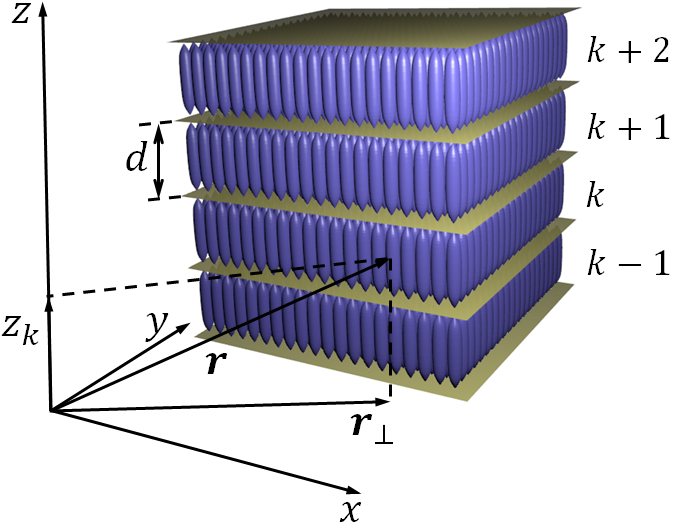}
	\caption{A schematic sketch of a LC in the smectic or hexatic phase. Parallel layers are formed by elongated LC molecules oriented perpendicular to the layer plane. The distance between the neighboring layers is $d$.  The $z$-axis is chosen to be perpendicular to molecular layers, while $\textbf{r}_\bot$ is parallel to the layer plane.
	}
	\label{3DLC}
\end{figure*}
\begin{figure*}
	\includegraphics[width=85mm]{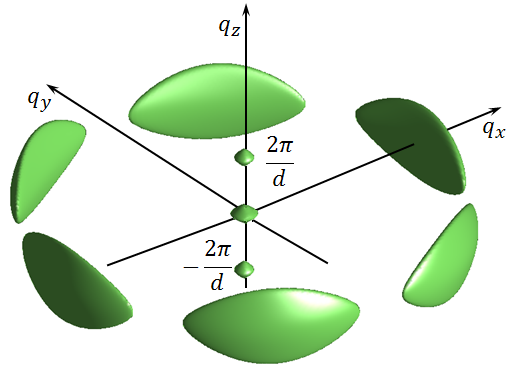}
	\caption{3D distribution of the scattered intensity $I(\textbf{q})$ for the thick multilayer LC in the hexatic phase close to the smectic-hexatic phase transition \cite{Oswald}. It consists from six large peaks corresponding to the scattering from a single LC layer and a set of Bragg peaks on the $q_z$-axis ($\textbf{q}_\bot=0$) that appears due to the presence of many parallel layers.
	}
	\label{StructureFactor}
\end{figure*}
Let us consider a three-dimensional (3D) liquid crystal (LC) film that can be described as a stack of $N$ parallel molecular layers separated by a distance $d$.
Let us introduce a coordinate system as it is shown in Fig. \ref{3DLC} in such a way that the $z$-axis is perpendicular to the layers.
The vector $\textbf{r}$ can be decomposed into two components, $\textbf{r}=\{\textbf{r}_\bot,z\}$, where $\textbf{r}_\bot$ lies in the plane parallel to molecular layers.
We assume that the electron density of the first layer can be described with the function $\rho_0(\textbf{r}_\bot,z)$.
We will enumerate layers with the index $k=0,...,N-1$.
We assume, that all layers have the same structure, so any layer can be obtained from the first one by translating the first layer by the vector $\{\textbf{r}_{\bot k},z_k\}$, where $z_k=kd$ is the separation distance in the vertical direction.
It means that all layers are oriented in the same way and there is complete angular correlation between all LC layers.
We will also assume that there are no positional correlations between the layers, the component $\textbf{r}_{\bot k}$ is some random number for each layer.
Thus, the total electron density of such a stack of molecular layers $\rho(\textbf{r}_\bot,z)$ can be written as a sum over all layers
\begin{equation*}
\label{Eq13}
\rho(\textbf{r}_\bot,z)=\sum_{k=0}^{N-1}\rho_0(\textbf{r}_\bot-\textbf{r}_{\bot k},z-z_k) \, ,
\end{equation*}
where $\textbf{r}_{\bot k}=0$ and $z_k=0$ for $k=0$.
The coherent x-ray scattering amplitude $A(\textbf{q})$ from such sample in kinematical approximation is the Fourier transform of the function $\rho(\textbf{r}_\bot,z)$
\begin{equation}
\label{Eq14}
A(\textbf{q})=\int \rho(\textbf{r}_\bot,z) e^{-i \textbf{q}_\bot \textbf{r}_\bot} e^{-i q_z z} \mathrm{d}\textbf{r}_\bot \mathrm{d}z
=A_0(\textbf{q}) \sum_{k=0}^{N-1} e^{-i \textbf{q}_\bot \textbf{r}_{\bot k}} e^{-i q_z kd}\,  ,
\end{equation}
where
\begin{equation}
\label{Eq14a}
A_0(\textbf{q})=\int \rho_0(\textbf{r}_\bot,z) e^{-i \textbf{q}_\bot \textbf{r}_\bot} e^{-i q_z z} \mathrm{d}\textbf{r}_\bot \mathrm{d}z
\end{equation}
is the Fourier transform of an electron density $\rho_0(\textbf{r}_\bot,z)$ of a single layer.
In the smectic phase there is no preferential orientation of intermolecular  bonds, so the intensity that would scatter from a single molecular layer, $I_0(\textbf{q})=|A_0(\textbf{q})|^2$, has cylindrical symmetry.
Within the $x-y$ plane $I_0(\textbf{q})$ has a form of concentric rings, that correspond to the average intermolecular distance.
In the $z$-direction the scattered intensity $I_0(\textbf{q})$ is determined by the molecular form factor.
In the hexatic phase the bond-orientationl order appears in each layer, so $I_0(\textbf{q})$ has a sixfold rotational symmetry around the $z$-axis and it consists of six separated peaks instead of a continuous ring in the smectic phase.

The scattered intensity from many LC layers is an averaged squared modulus of the amplitude $I(\textbf{q})=\langle
|A(\textbf{q})|^2\rangle$, where angular brackets denote ensemble averaging \cite{Chaikin}.
Using Eq. (\ref{Eq14}) we obtain for the scattering intensity from a stack of LC layers
%Thus, we can write the scattering intensity from a thick LC as a product of the intensity that would scatter from a single molecule layer $I_0(\textbf{q})$ and some scaling factor $\langle |S_N|^2 \rangle$
%
\begin{equation}
\label{Eq15}
I(\textbf{q})= I_0(\textbf{q}) \langle |S_N|^2 \rangle \, ,
\end{equation}
where
\begin{equation}
\label{Eq15a}
I_0(\textbf{q})= |A_0(\textbf{q})|^2
\end{equation}
and
\begin{equation}
\label{Eq16}
S_N =  \sum_{k=0}^{N-1} e^{-i \textbf{q}_\bot \textbf{r}_{\bot k}-i q_z kd}\, .
\end{equation}

Let us consider the case when $\textbf{q}_\bot\ne 0$.
We can write
\begin{equation*}
\label{Eq16a}
|S_N|^2=\sum_{k,k'} e^{-i\phi_{k,k'}-iq_z(k'-k)d}\, ,
\end{equation*}
where $\phi_{k,k'}=\textbf{q}_\bot(\textbf{r}_{\bot k'}-\textbf{r}_{\bot k})$ is a randome phase shift between the LC layers.
When averaged value $\langle |S_N|^2 \rangle$ is calculated one has to take into account the fact, that only exponent $e^{-i\phi_{k,k'}}$ has to be averaged, because all other factors do not depend on any random variable.
So we can write
\begin{equation}
\label{Eq16b}
\langle|S_N|^2\rangle=\langle\sum_{k,k'} e^{-i\phi_{k,k'}-iq_z(k'-k)d}\rangle=
\sum_{k,k'} \langle e^{-i\phi_{k,k'}} \rangle e^{-iq_z(k'-k)d}=N\, ,
\end{equation}
because for $k\ne k'$  the averaged value $\langle e^{-i\phi_{k,k'}} \rangle$ equals to zero, since phase $\phi_{k,k'}$ is random \cite{Goodman2,Goodman}, so only terms with $k=k'$ contribute to the sum (\ref{Eq16b}).
%Thus,  we can write $\langle |S_N|^2 \rangle=N$.
%Its mean amplitude equals to square root of the number of terms,$\sqrt{N^2}$, so $\langle |S_N|^2 \rangle=N$.

If $\textbf{q}_\bot=0$ then the factor $\langle |S_N|^2 \rangle$ in Eq. (\ref{Eq15}) can be rewritten without averaging, because in this case $S_N$ does not depend on any random variable
\begin{equation}
\label{Eq17a}
\langle |S_N|^2 \rangle_{\textbf{q}_\bot = 0}= \sum_{k,k'=0}^{N-1} e^{i q_z (k'-k)d}\xrightarrow{N\gg1}\Big(\frac{2\pi N}{d}\Big)^2\sum_{n}\delta(q_z-q_n) \, .
\end{equation}
Here we introduced $q_n=2\pi n /d$, where $n$ is an integer, that defines a reciprocal lattice to the 1D lattice that is formed by equidistant LC layers.
If there were some positional correlations between the layers, Bragg peaks would also appear for $\textbf{q}_\bot\ne0$.

Finally, we can combine the obtained results and write for the intensity scattered from the layered LC film
\begin{equation}
\label{Eq18}
I(\textbf{q}_\bot,q_z)= NI_0(\textbf{q})+\delta(\textbf{q}_\perp)\Big(\frac{2\pi N}{d}\Big)^2\sum_{n}\delta(q_z-q_n) \, .
\end{equation}
The first term in Eq. (\ref{Eq18}) is $N$ times the intensity scattered from a single molecular layer and the second term is proportional to $N^2$ and it represents a set of Bragg peaks that comes from a stack of parallel layers (see Fig. \ref{StructureFactor}).
%In the experiment described in the paper an x-ray beam was coming perpendicular to LC layers, so from all Bragg peaks at $q_z=q_n$ the Ewald sphere was cutting only the Bragg peak at $q_z=0$ that was covered by a beamstop.
%So for our experimental conditions we can neglect the sum in Eq. (\ref{Eq18}), since we were not measuring this part of reciprocal space in the described geometry of the experiment, and simply write $I(\textbf{q})=NI_0(\textbf{q})$.

\section{The projection-slice theorem}
\label{AppC}
\begin{figure*}
	\includegraphics[width=65mm]{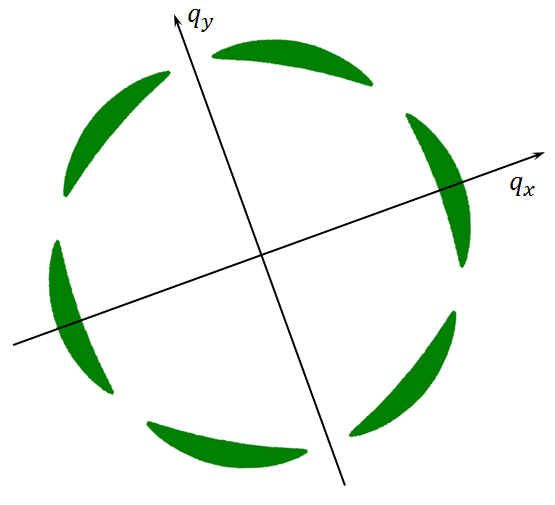}
	\caption{The cross section $q_z=0$ of the hexatic structure factor $S(\textbf{q}_\bot,q_z)$ shown in Fig. \ref{StructureFactor}. $x$-axis is chosen to go through the center of one diffraction peak.
	}
	\label{Axis}
\end{figure*}

In two dimensions the projection-slice theorem states, that the results of the following two calculations are equal \cite{Poularikas}:
\begin{itemize}
	\item Take a two-dimensional function $g(\textbf{r})$, project it onto a (one-dimensional) line, and do a Fourier transform of that projection.
	\item Take the same function $g(\textbf{r})$, do a two-dimensional Fourier transform first, and then slice it through its origin, which is parallel to the projection line.
\end{itemize}
The functions $S(\textbf{q})-1$ and $g(\textbf{r})-1$ are connected to each other by a two-dimensional Fourier transform (see Eq. (\ref{Eq1})).
Let us assume that the structure factor can be described by the Lorentzian function $S(q)=\gamma^2/((q-q_0)^2+\gamma^2)$ along some direction, that we will denote as $q_x$-axis of a Cartesian coordinate system as it is shown in Fig. \ref{Axis}.
The half width at half maximum of the Lorentzian-shaped peak of the structure is determined by parameter $\gamma$, that is related to correlation length of the system by $\xi=1/\gamma$.
Since the structure factor is symmetric, i.e. $S(-\textbf{q})=S(\textbf{q})$ (Fig. \ref{Axis}), the slice of $S(\textbf{q})$ in this direction $q_y=0$ will be described as a sum of two Lorentzian functions
\begin{equation}
\label{Eq30}
S(q_x,0)=\frac{\gamma^{2}}{(q_x-q_0)^2+\gamma^2}+\frac{\gamma^{2}}{(q_x+q_0)^2+\gamma^2} \, .
\end{equation}

\begin{figure*}
	\includegraphics[width=65mm]{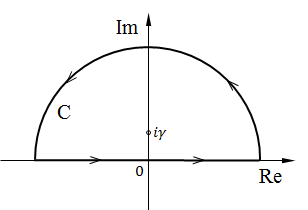}
	\caption{The contour $C$ of the integral (\ref{Eq30a}) in the complex plane.
	}
	\label{contour}
\end{figure*}

The projection-slice theorem allows one to determine the value of the parameter $\gamma$ and corresponding correlation length $\xi=1/\gamma$ through the projection of the real-space function $g(\textbf{r})-1$ on the $x$-axis.
In order to do this one has to calculate a Fourier transform of the function $S(q_x,0)$

\begin{equation}
\label{Eq30a}
\int\limits_{-\infty}^{+\infty}S(q_x,0)e^{-iq_x x}\mathrm{d}q_x=\gamma^2(e^{iq_0 x}+e^{-iq_0 x})\int\limits_{-\infty}^{+\infty} \frac{e^{-iq_x x}}{q_x^2+\gamma^2}\mathrm{d}q_x
=2\gamma^2\cos{(q_0x)}\oint\limits_{C} \frac{e^{-iq_x x}}{q_x^2+\gamma^2}\mathrm{d}q_x \, .
\end{equation}
Here we closed the integration contour $C$ with a semicircle of infinite radius as it is shown in Fig. \ref{contour}, because the integral along the semicircle is infinitesimal.
This integral can be conveniently calculated by Cauchy's residue theorem \cite{Ahlfors} that states that integral over the contour $C$ of an analytical  function $f(z)$ can be calculated as a sum
\begin{equation}
\label{Eq30b}
\oint\limits_{C}f(z)\mathrm{d}z=2\pi i\sum_{z_j}\operatorname{Res}(f(z),z_j) \, ,
\end{equation}
where $z_j$ are the poles of the function $f(z)$ inside the contour $C$ and $\operatorname{Res}(f(z),z_j)$ denotes the residue of $f(z)$ in $z_j$.
The function $e^{-iq_x x}/(q_x^2+\gamma^2)$ has only one pole $z=i\gamma$ in the  upper half-plane, so the integral (\ref{Eq30a}) can be easily evaluated and one finally obtains
\begin{equation}
\label{Eq30c}
\int\limits_{-\infty}^{+\infty}S(q_x,0)e^{-iq_x x}\mathrm{d}q_x=2\gamma^2 \cos(q_0x)\times 2\pi i\operatorname{Res}\Big( \frac{e^{-iq x}}{q^2+\gamma^2}, q=i\gamma \Big)=2\pi\gamma\cos{(q_0x)} e^{-\gamma |x|} \, .
\end{equation}

Finally, we can write a Fourier transform of a slice of the function $S(\textbf{q})-1$:
\begin{equation}
\label{Eq31}
\int_{-\infty}^{+\infty}\big(S(q_x,0)-1\big)e^{-iq_x x}\mathrm{d}q_x=2\pi\gamma\cos{(q_0x)} e^{-\gamma |x|}-2\pi\delta(x) \, .
\end{equation}
According to the projection slice theorem we would obtain the same result by calculating the projection of the function $g(\textbf{r})-1$ on the $x$-axis.
Thus, for any $x\ne0$, we can introduce the function $g_{\text{pr}}(x)$ that is the projection of the function $g(\textbf{r})-1$ onto the direction $y=0$,
\begin{equation}
\label{Eq32}
g_{\text{pr}}(x)\equiv\int_{-\infty}^{+\infty}\big(g(x,y)-1\big)\mathrm{d}y=A\cos{(q_0x)} e^{-\gamma |x|} \, ,
\end{equation}
where $A=2\pi\gamma/\langle \rho \rangle$.

\section{PDF of a system with a bond-orientational order}
\label{AppD}
Let us consider the structure factor of a 2D system with bond-orientational order to be represented as a product
\begin{equation}
\label{Eq19}
S(q,\phi)=S(q)S(\phi) \, ,
\end{equation}
where $S(q)$ describes the positional order and $S(\phi)$ corresponds to the bond-orientational order in the system.
We will consider only the first-order diffraction peaks, because the higher order peaks have weaker intensity.
For the systems with short-range positional order the structure factor $S(q)$ can be described by the Lorentzian function
\begin{equation}
\label{Eq20}
S(q)=\frac{\gamma^{2}}{(q-q_0)^2+\gamma^2} \, ,
\end{equation}
where $q_0$ determines the scattering intensity maximum position and $\gamma$ determines half width at half maximum (HWHM) of the peak.
%For large values of $q$ the structure factor should asymptotically approach unity, that does not taken into account in the Eq. (\ref{Eq20}).
The angular dependence of the structure factor $S(\phi)$ can be represented as a Fourier series
\begin{equation}
\label{Eq21}
S(\phi)=\sum\limits_{n=-\infty}^{+\infty}C_{n}\exp{(in\phi)} \, ,
\end{equation}
where the coefficients $C_{n}$ describe the shape of the diffraction peaks in the azimuthal direction.
%\footnote{For example,  in case of sixfold rotational symmetry $C_{6m}=\sqrt{18\sigma^2/\pi}\exp{(-(6m)^2\sigma^2/2)}$ for the Gaussian peaks with width $\sigma$, if  maximum of one of six peaks is on the reference axis.}.
In the case of a mono-domain system one can set all coefficients $C_{n}$ to be real by proper choice of the reference axis.

The angular Fourier components of the structure factor given by Eqs. (\ref{Eq19}-\ref{Eq21}) are
\begin{equation}
\label{Eq22}
S_{n}(q)=S(q)C_{n} \,  .
\end{equation}
According to the Eq. (\ref{Eq12}) the corresponding Fourier coefficients of the PDF $g_n(r)$ are
\begin{equation}
\label{Eq23}
g_{n}(r)=\delta_{0,n}+\frac{1}{2\pi \langle \rho \rangle i^{n} }\int\limits_{0}^{+\infty}\Big(\frac{\gamma^2C_{n}}{(q-q_0)^2+\gamma^2}-\delta_{0,n}\Big)J_{n}(qr)q\mathrm{d}q \, .
\end{equation}
Substituting the FCs (\ref{Eq23}) into expression for the PDF (\ref{Eq2}) we obtain
\begin{equation}
\label{Eq24}
g(r,\theta)=1+
\frac{1}{2\pi \langle \rho \rangle}
\sum_{n=-\infty}^{+\infty}	
(-i)^ne^{in\theta} \int\limits_{0}^{+\infty}\Big(\frac{\gamma^2C_{n}}{(q-q_0)^2+\gamma^2}-\delta_{0,n}\Big)J_{n}(qr)q\mathrm{d}q \, .
\end{equation}
Here angular and radial parts of the PDF appear to be coupled to each other, although the structure factor (\ref{Eq19}) was taken as a product of angular and radial parts.
We will show now, that for large distances the PDF $g(r,\theta)$ can be also represented as a product of angular and radial parts.

The integral $\int\limits_{0}^{+\infty}J_0(qr)qdq$ that appears in Eq. (\ref{Eq24}) from the term with delta function for $n=0$ decays as $r^{-\frac{3}{2}}$ for large distances.
We will neglect this integral, because as we will show below, all other therms decay slower, namely as $r^{-\frac{1}{2}}$, so one can write for the PDF
\begin{equation}
\label{Eq24a}
g(r,\theta)\approx1+
\frac{\gamma^2}{2\pi \langle \rho \rangle}
\sum_{n=-\infty}^{+\infty}	
(-i)^nC_{n}e^{in\theta} \int\limits_{0}^{+\infty}\frac{1}{(q-q_0)^2+\gamma^2}J_{n}(qr)q\mathrm{d}q \, .
\end{equation}
If the width of the Lorentzian function $\gamma$ is small in comparison with a characteristic value $q_0$ ($0<\gamma\ll q_0$) the integrand is small for all values of $q$ except the vicinity of the point $q=q_0$, so the main contribution to this integral comes from the point $q=q_0$.
It means that the integration interval can be formally extended from $[0, +\infty)$ to $(-\infty, +\infty)$.
If $r$ is large enough ($q_0r\gg n^2$) the Bessel function $J_{n}(qr)$ can be replaced with its asymptotic expression \cite{Watson}
\begin{equation}
\label{Eq25}
J_{n}(qr)\approx\sqrt{\frac{2}{\pi qr}}\cos{\Big( qr-\frac{\pi n}{2} -\frac{\pi}{4}\Big)}\, .
\end{equation}
Finally, one can write for the integral in Eq. (\ref{Eq24a})
\begin{multline}
\label{Eq26}
\int\limits_{0}^{+\infty}\frac{1}{(q-q_0)^2+\gamma^2}J_{n}(qr)q\mathrm{d}q\approx
\sqrt{\frac{2}{\pi r}}\int\limits_{-\infty}^{+\infty}\frac{\sqrt{q} \cos{\Big( qr-\frac{\pi n}{2}-\frac{\pi}{4}\Big)}} {(q-q_0)^2+\gamma^2}\mathrm{d}q=\\
=\sqrt{\frac{2}{\pi r}}
\operatorname{Re}
\Bigg\{
\int\limits_{-\infty}^{+\infty}
\frac{\sqrt{q} \exp{\big(iqr-i\frac{\pi n}{2}-i\frac{\pi}{4}\big)}} {(q-q_0)^2+\gamma^2}\mathrm{d}q
\Bigg\} \, ,
\end{multline}
where $\operatorname{Re}\{z\}$ denotes the real part of a complex number $z$.

The final integral in this equation can be calculated using Cauchy's residue theorem if one closes the integration contour $C$ with a semicircle of infinite radius, exactly as it was done in the calculation of the integral (\ref{Eq30a}).
In this case the integrand has a simple pole $q=q_0+i\gamma$ in the upper half-plane, so one can write
\begin{multline}
\label{Eq27}
\int\limits_{-\infty}^{+\infty}\frac{\sqrt{q} \exp{\big(iqr-i\frac{\pi n}{2}-i\frac{\pi}{4}\big)}} {(q-q_0)^2+\gamma^2}\mathrm{d}q=
e^{-i\frac{\pi n}{2}-i\frac{\pi}{4}}\oint\limits_{C}\frac{\sqrt{q} e^{iqr}} {(q-q_0)^2+\gamma^2}\mathrm{d}q=\\
=e^{-i\frac{\pi n}{2}-i\frac{\pi}{4}} 2\pi i \operatorname{Res}\Big( \frac{\sqrt{q} e^{iqr}} {(q-q_0)^2+\gamma^2}, q=q_0+i\gamma \Big)
= \frac{\pi}{\gamma}\sqrt{q_0+i\gamma}\exp{\Big( iq_0r-i\frac{\pi n}{2}-i\frac{\pi}{4} \Big)} \, .
\end{multline}

We can neglect the term $i\gamma$ in comparison to $q_0$ and then calculate the integral (\ref{Eq26})
\begin{equation*}
\label{Eq28}
\int\limits_{0}^{+\infty}\frac{1}{(q-q_0)^2+\gamma^2}J_{n}(qr)q\mathrm{d}q\approx
\sqrt{\frac{2 \pi q_0}{\gamma}}\frac{e^{-\gamma r}}{\sqrt{\gamma r}}\cos{\big( q_0r-\frac{\pi n}{2}-\frac{\pi}{4}\big)} \, .
\end{equation*}

Finally, under the assumptions $q_0r\gg(6m)^2$ and $\gamma\ll q_0$ that we used above, we can write the expression for the PDF $g(r,\theta)$,
\begin{equation}
\label{Eq29}
g(r,\theta)\longrightarrow 1
+\frac{\gamma}{\langle \rho \rangle} \sqrt{\frac{q_0}{2\pi}} \frac{e^{-\gamma r}}{\sqrt{r}}\times\sum_{n=-\infty}^{+\infty}(-i)^n e^{in\theta}C_n\cos{\big( q_0r-\frac{\pi n}{2}-\frac{\pi}{4}\big)}\, .
\end{equation}
Having in mind that due to the symmetry of a diffraction pattern the coefficients $C_n$ have non-zero values only for even $n$, one can note that the factor $(-i)^n$ and phase shift of $-\pi n/2$ in the cosine function exactly compensate each other.
Thus, one can write the asymptotic expression for the PDF $g(r,\theta)$ for a 2D system with bond-orientational order
\begin{equation}
\label{Eq29}
g(r,\theta)\longrightarrow 1
+B \frac{e^{-\gamma r}}{\sqrt{r}} \cos{\big( q_0r-\frac{\pi}{4}\big)} \times\sum_{n=-\infty}^{+\infty} C_n e^{in\theta}\, ,
\end{equation}
where $B=\gamma/\langle \rho \rangle \sqrt{q_0/2\pi}$.
This formula means that the magnitude of oscillations of the PDF $g(r,\theta)$ at large distances decays as $e^{-\gamma r}/\sqrt{r}$ and the angular dependence of the PDF $g(r,\theta)$ is the same as the angular dependence of the structure factor $S(\phi)$ in Eq. (\ref{Eq21}).

\bibliography{References_PDF_ARXIV}	

\begin{thebibliography}{34}
\bibitem [{Str()}]{StructureFactorNote}%
  \BibitemOpen
  \href@noop {} {}\bibinfo {note} {The measured LC film (about $5\;\mu\text{m}$
  thick) consists of a stack of many molecular layers and strictly speaking can
  not be considered as a pure 2D system. However, one can show that for such
  layered system in the absence of positional correlations between the layers
  the structure factor is the same as for a single layer (see Appendix
  \ref{AppB}).}\BibitemShut {Stop}%
\bibitem [{Neg()}]{NegativePDF}%
  \BibitemOpen
  \href@noop {} {}\bibinfo {note} {The PDF has negative values close to the
  point $r=0$, that should be considered as an artefact caused by the finite
  experimental resolution.}\BibitemShut {Stop}%
\end{thebibliography}
\end{document}